REVIEW

# Tracing protein and proteome history with chronologies and networks: folding recapitulates evolution

Gustavo Caetano-Anollés[1,2], M. Fayez Aziz[1], Fizza Mughal[1] and Derek Caetano-Anollés[3]

[1]Evolutionary Bioinformatics Laboratory, Department of Crop Sciences, Urbana, Illinois, USA; [2]C. R. Woese Institute for Genomic Biology, University of Illinois, Urbana, Illinois, USA; and [3]Broad Institute of MIT and Harvard, Cambridge, Massachusetts, USA

CONTACT  Gustavo Caetano-Anollés  gca@illinois.edu  Evolutionary Bioinformatics Laboratory, Department of Crop Sciences, University of Illinois, Urbana 61801 Illinois, USA

**ABSTRACT**

**Introduction:** While the origin and evolution of proteins remain mysterious, advances in evolutionary genomics and systems biology are facilitating the historical exploration of the structure, function and organization of proteins and proteomes. Molecular chronologies are series of time events describing the history of biological systems and subsystems and the rise of biological innovations. Together with time-varying networks, these chronologies provide a window into the past.

**Areas covered:** Here, we review molecular chronologies and networks built with modern methods of phylogeny reconstruction. We discuss how chronologies of structural domain families uncover the explosive emergence of metabolism, the late rise of translation, the co-evolution of ribosomal proteins and rRNA, and the late development of the ribosomal exit tunnel; events that coincided with a tendency to shorten folding time. Evolving networks described the early emergence of domains and a late 'big bang' of domain combinations.

**Expert opinion:** Two processes, folding and recruitment appear central to the evolutionary progression. The former increases protein persistence. The later fosters diversity. Chronologically, protein evolution mirrors folding by combining supersecondary structures into domains, developing translation machinery to facilitate folding speed and stability, and enhancing structural complexity by establishing long-distance interactions in novel structural and architectural designs.

**Article highlights**

- Advances in evolutionary genomics and systems biology enable the historical exploration of the structure, function and organization of proteins and proteomes.
- Proteins fold cooperatively into many layers of molecular organization through asynchronous formation of secondary structures, their co-translational stabilization along the ribosomal exit tunnel, and long-distance interactions that guide the free-energy landscape towards the native state.
- Proteins also embed autonomous and highly reused folding elements, including supersecondary structures and structural domains, which behave as structural, functional and evolutionary units.
- The cornerstone of protein hierarchical classification is homology of structural domains, the existence of shared-and-derived features in the sequence, structure and function of these modules.
- Comparative genomic analyses of domain distributions in the proteomes of superkingdoms and viruses reveal a common core suggestive of ancient ancestors and a periphery describing a tripartite cellular world and the very early origin of viruses and Archaea.
- Phylogenomic analyses permits construction of molecular chronologies and evolving networks, which can be linked to the geological record using a molecular clock of folds.
- Chronologies of domain structures and architectures uncover the early evolutionary rise of sandwich, bundle, barrel and more complex structures, in that order, and a late 'big bang' of domain combinations.
- Chronologies of domain families reveal the explosive emergence of metabolism, the late rise of translation, the co-evolution of ribosomal proteins and rRNA, and the late development of the exit tunnel, events that coincided with a tendency to shorten folding time.
- Evolution of protein structure mirrors the folding process by combining supersecondary structures into domains, developing complex translation machinery that facilitate folding speed and stability, and enhancing structural complexity by establishing long-distance interactions in novel structural and architectural designs

# 1. Introduction

A proteome represents the entire collection of proteins that is encoded in the genome of an organism. This potential set of gene products can be either inferred from genome sequences or can be experimentally determined [1]. For example, a genomic catalog of thousands of deep RNA sequencing experiments of the human genome revealed a total of ~320,000 transcripts encoded by ~42,000 genes, ~20,000 of which are translated into proteins [2]. However, not all genes of a genome may be expressed by a cell, tissue or organism at a given time and at a certain environmental condition. A mass spectrometric analysis of protein levels encoded by over 12,000 human genes across 201 samples of 32 normal human tissues revealed that ~85% proteins were present in all tissues [3]. The relative abundance of these common proteins together with presence/absence of a minority protein set helped explain biological processes that require the interplay of multiple tissues. Tissue-specific distribution of enzymes revealed a coordinated control of metabolic reactions, while tissue-enriched proteins provided insights into phenotypes of genetic diseases. Mass spectrometric measurements allow visualization of proteomes with unprecedented quantitative detail, which can be recalibrated to estimate the number of proteins per unit cell volume [4]. About 2-4 million proteins per cubic micron populate the cells of varied organisms. Typical bacteria will contain about 3 million proteins (e.g. *Escherichia coli*) while a typical mammalian cell will contain over a billion, four orders of magnitude more. This reality highlights how crowded is the molecular environment of the cell.

Most proteins are both structured and intrinsically flexible. Their polypeptide chains fold into compact atomic three-dimensional (3D) arrangements that are organized around structural, functional and evolutionary *modules*. These modules are recurrently present in various molecular contexts. The human genome, for example, embodies ~1,500 distinct combinations of folded structures [5]. Conversely, a substantial number of proteins lack typical structure. They represent intrinsically disordered proteins, molecules that lack significant constraints on internal degrees of freedom of the polypeptide chain. *Intrinsic disorder* is also present in structured proteins in the form of intrinsically disordered regions [6]. These regions exhibit highly dynamic conformations that resemble either random-coils, molten globules or flexible linkers. Disordered regions are often evolutionarily conserved and needed for molecular recognition, regulation and signaling (e.g. [7]). Significant surveys of protein modules and intrinsic disorder with advanced computational methods have for example generated protein taxonomies for the classification of the protein world. Similarly, the distribution of protein structure and intrinsic disorder in organisms provide the necessary tools to understand the origin and evolution of proteomes.

Here we explore how developments in evolutionary genomics and systems biology are helping understand the origin and evolution of proteins and proteomes. We start by addressing protein structural complexity, over what has been already reviewed [8,9]. Proteins are highly organized entities that fold into structures that exhibit many levels (layers) of structural organization. These structures are highly diverse and have been the subject of exhaustive classification. We focus on chronologies that describe how protein modules are becoming structured in the protein world and how they originate and spread in evolving and increasingly complex proteomes. Our emphasis is to prompt critical thinking that could help untangle processes of molecular emergence and diversification that are operating in our planet.

## 2. The protein makeup of proteomes

### 2.1. Proteins are functional nanoparticles

Proteins are biological macromolecules with properties of nanoparticles. Typical 5-500 kDa proteins have diameters that range 2-10 nm [10]. They are made up of one or more relatively long polypeptides that fold into globular, fibrous or membrane forms. Their name [from the Greek πρώτειος (prōteios) and πρῶτος (prōtos), meaning "primary" and "first"] suggests their very primordial origin, which as we will discuss below makes justice to their very early origin. A polypeptide is a single linear heteropolymer chain of amino acids covalently bonded together by peptide bonds that link the carboxyl and amino groups of adjacent amino acid residues. Proteins synthesized by the ribosomal machinery of the cell typically contain polypeptides of more than 20 amino acid residues in length. Shorter molecules are generally synthesized by the non-ribosomal protein synthetase (NRPS) machinery and are called 'peptides'. The typical polypeptide lengths of proteins range from tens of amino acids to thousands [11-13], with mean values of 329, 365 and 532 amino acids for proteomes of the Archaea, Bacteria and Eukarya superkingdoms, respectively [14]. Small proteins have been overlooked in genome annotations. The shortest described so far is the 11-amino acid Tal protein that influences *Drosophila* development, which contrasts with the largest, the 34,350-amino acid sarcomeric titin protein [15]. Protein monomers belong to the standard complement of 20 L-α-amino acids (plus the atypical selenocysteine and pyrrolysine), which provides variable side chains to a central protein backbone and a multiplicity of physicochemical properties (e.g. size, polarity, charge, hydropathy) for the stability and function of the macromolecules. This relatively small amino acid repertoire contrasts with the over 500 monomer types used by non-ribosomal peptides synthesized by NRPSs [16]. Besides canonical amino acids, this diverse monomeric pool includes β-amino acids, methylated, hydroxylated, and D-configured proteogenic amino acids, and even fatty acid building blocks.

Proteins hold the biological functions of the cell, sometimes by associating together to form stable protein complexes. They define the catalytic activities and transport processes that are characteristic of the enzymes of metabolic pathways. They mediate numerous processes including gene expression, membrane transport, cell signaling and adhesion, immune responses, and the cell cycle. They act as cellular scaffolds and provide structural and mechanical functions. Proteins control all levels of a hierarchy that starts with the gene and defines structure and function of the cell, tissue, organ and organism [17]. The Gene Ontology (GO) consortium now provides a systematic view of the molecular function (*mf*), biological process (*bp*) and cellular compartment (*cc*) annotations of their biological activities [18]. Protein functions are often mediated by binding sites, regions of the macromolecules than specifically bind to small and large ligands. A recent census of atomic interactions between proteins and small ligands surveyed the frequency, geometry and impact of a multiplicity of interaction types, recapitulating well-known rules driving ligand binding and uncovering overlooked types [19]. Many databases that annotate protein-protein interactions exist that help dissect molecular functions that involve experimental and predicted interactions between proteins (e.g., [20]). A very challenging focus is the existence of catalytic site mechanisms and allosteric regulatory interactions. An analysis of catalytic residues annotated in the Mechanism and Catalytic Site Atlas (M-CSA) revealed catalytic residues were highly conserved and were restricted to a small set performing a limited number of catalytic functions [21]. Some specificity rules could be derived for certain functions by identifying chemical groups acted upon by catalytic residues. This kind of information is necessary to understand how mutations impact both enzyme function and evolution. Allosteric interactions control protein activities when effector molecules bind to sites distant from the active (usually catalytic) sites [22]. A very recent experimental-computational effort dissected the genotype-phenotype landscape of an allosteric protein, the LacI repressor protein from *Escherichia coli* [23]. A quantitation of dose-response curves for nearly $10^5$ variants of the genetic sensor matched neural network-model predictions but also revealed effects from combinations of nearly silent amino acid substitutions. Remarkably, active sites control substitutions in a network spanning over 80% of the residues of an enzyme, with conservation decreasing approximately linearly with increasing distance [24].

## 2.2. Protein folding is central to protein function

The processual foundation of protein structure and its many organizational levels is the folding of the linear polypeptide chain into an 'origami' of 3D atomic arrangements of the protein backbone. These folded structures accomplish several goals including conformation energy minimization of residues in the chain, maximization of hydrogen-bonding of polar groups, and burying of hydrophobic residues in well packed 3D structures. Protein folding can be described by a firmly established 'funneled energy landscape' in which protein sequences quickly fold to their native conformations following a principle of minimal frustration [25]. The energy landscape paradigm implies that proteins belonging to a same fold family (with common functions and structures) should have similar folding rates regardless of their time of evolutionary origin, while differences in thermodynamic stability should manifest through distinct unfolding rates. Remarkably, extant and resurrected thioredoxin proteins exhibiting a wide spectrum of thermodynamic stabilities folded with similar rates while rates of unfolding of thioredoxin variants differed over 3 orders of magnitude, directly supporting the minimum frustration principle [26]. We note however that phylogenomic and structural analyses revealed that folding speed of individual fold families increased in a chronology between 3.8 and 1.5 billion years ago (Gya) [27]. This suggests that the universe of protein structure was initially shaped by an evolutionary optimization for rapid folding, a tendency that reverts starting 1.5 Gya following the rise of multidomain proteins. Folding occurs cooperatively through asynchronous formation of numerous folding intermediates, starting with small-sized (~20 residues) 'foldon' building blocks that repeatedly fold and unfold into short helices, turns and bends [28]. These emergent elements of secondary structure are soon complemented with 'long distance' interactions that form strands, bridges and sheets, which guide the free energy landscape through stepwise pathways toward the native state. Advanced pulse labeling mass spectrometry experiments support these cooperative foldon interactions [29]. In addition, the folding of proteins involves co-translational regulation in the ribosome (reviewed by [30]). Folding is vectorially initiated within the ribosomal exit tunnel, a tube-like structure ~80-100 Å long and ~10-20 Å wide, by entropically stabilizing the formation of α-helices, hairpins, and in some cases, small α-helical domains. Nascent folding and stabilization of folding intermediates has been extensively confirmed by FRET, PET, time-resolved translation assays, and cryo-EM experiments (e.g. [31]). The chaperoning effects of ribosomal interactions increase folding efficiency but depend on the rate of biosynthesis of the nascent polypeptide chain during translational elongation and the harmonized usage of synonymous codons [32]. Covalent modifications such as N-nitrosylation and S-gluthathionylation of cysteine residues that promote the formation of non-native disulfide bridges also occur deep inside the ribosomal exit tunnel prior to nascent chain release [33]. As partially folded polypeptide chains reach the exit tunnel vestibule, proteins acquire tertiary structure and then start to assemble into quaternary complexes [34]. There is still need to understand the role of the ribosomal surface in this process and how length-dependent free energy landscapes modulate co-translational folding within the confines of translation kinetics and the effect of molecular chaperones. Remarkably, the width of the exit tunnel narrows in the middle by a constriction from loops of ribosomal proteins (r-proteins) L4 and L22, which dissect the tunnel into an evolutionarily conserved upper part and a less-conserved lower part that ends in the tunnel vestibule [35]. Eukaryotic and archaeal lower tunnels were narrower than those of bacterial counterparts, a fact that will remain relevant in our discussion.

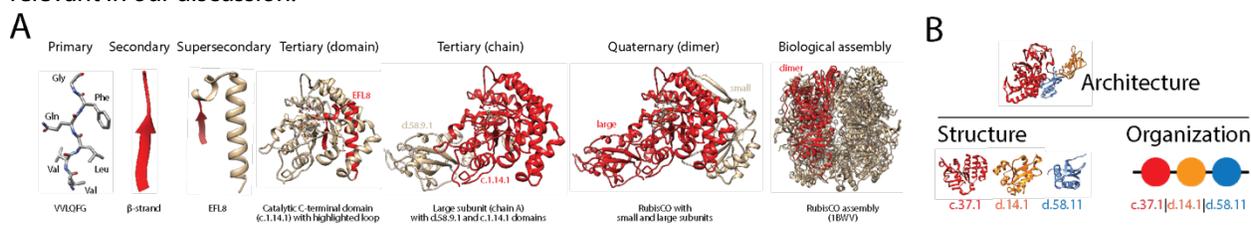

**Figure 1.** The makeup of proteins. A. Levels of molecular organization embedded in a protein biological assembly, the ribulose-1,5-bisphosphae carboxylase enzyme (RubisCO; PDB entry 1WDD). B. Elongation factor G from *Thermus thermophilus* (1DAR) illustrates how a domain architecture describes both the structure and organization of its structural domain components.

## 2.3. Proteins are organized entities

Proteins are highly diverse and embed many levels of molecular organization (Figure 1A). At the lowest organization levels, *primary structure* specifies the sequence of the monomeric subunits of the polymer and *secondary structure* determines the local folding of the polypeptide chain into regular structured segments. Secondary structure elements include helices (mostly α-helices) and extended chain segments called β-strands. Helical regions locally distort the protein backbone through close-range hydrogen-bonding electrostatic interactions established between main-chain carbonyl (C=O) and amine (N-H) groups typically located 3-5 amino acid residues away in the helical turn. In contrast, β-strands establish hydrogen-bonding interactions between the main-chain groups distantly located in the polypeptide chain, forming β-sheets in parallel and antiparallel arrangements that often curve into open and closed barrel structures. The 'helical' and 'sheet' components are separated by rigid stretches of the backbone with no standard secondary structure. These non-regular loop regions [36] generally harbor shorter turns and coils that change the direction of the polypeptide chain in space and are often of functional importance (e.g. β-turns [37]). Remarkably, α-helices can accumulate more mutations than β-strands without changing their structure, while at the same time diverging more rapidly in sequence than strands of a same structural domain [38]. Both helices and strands are however more robust to mutational change than regions with no secondary structure, making them more designable. These findings support the suggestion that protein folds resist mutational destabilization by increasing the density of contacts between residues [39].

Loops define a diverse group of building blocks made of helix, strand and turn segments that are generally ~25 to 30 amino acid residues long [40,41]. Examples include αα-hairpins, ββ-hairpins, and βαβ-elements. These *supersecondary structures* (Figure 1A) often collapse into extended flexible or rigid loop-shaped conformations stabilized by van der Waals locks [42,43]. The connector regions of these 'loop-and-lock' structures can be exposed to the protein surface and can interact with solvent, ligands, and macromolecules. They hold answers to protein function [44] and are known determinants of macromolecular dynamics and allostery [41]. While their biophysical properties may be governed by yet-to-be discovered evolutionary drivers, some recent studies identified combinable [45-48] and non-combinable [49] loop motif prototypes that were evolutionarily conserved and were likely responsible for the early rise of molecular functions in protein evolution. The first group of loop prototypes combine with others to form active sites that bind cofactors and exert molecular functions [47]. These 'elementary functional loops' (EFLs) were obtained by iterative derivation of sequence profiles from protein coding sequences in complete proteomes (a procedure that resembles PSI-BLAST) using a scoring function that weights profile positions according to their information content [45]. Redundancies from overlaps of steps of 10 residues were later removed by profile clustering using a distance measure that accounts for all possible profile-profile alignments. Network visualization of putative evolutionary connections between EFLs revealed striking patterns of motif reuse in archaeal proteins [46]. A number of network clusters unified (i) nucleotide/phosphate cofactor binding domains in ABC transporters, aminoacyl-tRNA synthetases, and helicases, (ii) metal cofactor binding domains, (iii) methylases and methyltransferases, (iv) transcriptional regulators, and (v) structural repeat building blocks typical of cell surface proteins. EFLs can be interpreted as ancient structural scaffolds that emerged in protein evolution from shorter protein modules typically 6-9 residues long [50,51] that were recruited to build higher levels of protein organization. Short modules that are omnipresent include 'aleph' loops and 'beth' sequences present in ancient ABC cassettes and other regions of ABC transporter proteins. In contrast with EFLs, highly repeated non-combinable loop structures that are present in popular folds appear to represent remnants of an ancient peptide 'vocabulary' that combined and accreted to form folded proteins during a primordial RNA-peptide world [49]. The comparative approach that was used to identify these supersecondary structures explored local sequence similarities with HHsearch and local structural similarities with TM-align in a set of protein structural domains unified by <30% sequence similarities. The machine learning tools revealed biphasic patterns in probability distributions that highlighted high-scoring subdomain-sized fragments. These were clustered in 65 groups, reduced to 40 groups to avoid redundancies and artifacts from domain boundaries, and aligned into folded loop structures that were 9-39 residue long. Loop structures that were identified embedded nucleic acid binding fragments with the helix-turn-helix and helix-hairpin-helix motifs and ribosomal protein motifs. Another group of loops was enriched in catalytic (P-loop motifs and dinucleotide-binding β-α-β motifs) and binding functions (coordination with metal ions and iron-sulfur clusters). A third general type of widely reused supersecondary structural motif has been described with lengths that vary from 35 to 200 amino acids [52,53]. They have been used to build networks of domains and motifs linked by motif reuse in domains [52]. Operationally, these motifs are selected from meaningful all-versus-all SSM or HHsearch alignments. In a more advanced implementation, candidate motifs are identified using an efficient dynamic programming algorithm and a two-step function that finds widely reused contiguous non-overlapping segments of widely different lengths, which are named 'themes' [53]. Interestingly, reuse increases with decreasing theme length following a power law, indicating a significant biased distribution of these motifs in proteins. In all of these approaches, (i) supersecondary motifs are identified by sequence and/or structure similarities, which may not necessarily carry evolutionary relationship; (ii) motif recurrence across proteins is considered driven by biological function; and (iii) motifs form complex patterns in proteins suggestive of an interplay of divergent vs. convergent evolution perhaps driven by the existence of rearrangements, duplications, and divergences. EFLs, loops and themes are therefore considered ancient evolutionary building blocks.

At higher levels of organization, striking regularities exist in how secondary structures assemble into tightly packed layered arrangements of the polypeptide chain [54]. These regularities in connectivities and relative orientation of secondary structures (topologies and architectures, respectively), which result in part from physical and chemical properties that are intrinsic [55,56], are responsible for protein *tertiary structure*. Tertiary structure first manifests in the formation of autonomous folding elements known as protein *structural domains* (Figure 1A). Domains are structural [57], evolutionary [58] and functional [59] units of proteins, mainly because of their 'compact globular' folded structure [first observed by Phillips [60] in lysozyme], their high evolutionary conservation [61] and their recurrent association with molecular functions [62]. Since proteins often contain more than one domain, domains appear repeatedly, individually or combined with other domains sometimes in unusually complex arrangements [5]. This enhances domain recurrence. Tertiary structure also manifests in supradomains, domain combinations that behave as modules and appear recurrently in multidomain proteins [63].

The *quaternary structure* of a protein involves self-assembly of different polypeptide chains into a multimolecular complex [64]. About 30-50% of proteins self-assemble into symmetrical complexes, in which multiple copies of a protein unit (homomer) interact with each other [65]. These assemblies of self-interacting units are invariably essential to the function of a protein. The ribulose-1,5-bisphosphae carboxylase (RubisCO), the enzyme responsible for $CO_2$ fixation in photosynthesis (Figure 1A), serves as an example. Uncovering the mutational landscape of residues of the quaternary complex revealed that in evolution, the acquisition of enhanced enzymatic activity was induced by mutations in catalytic sites and at the interface of the domains and subunits of RubisCO [66]. These mutations enhanced conformation flexibility of the catalytic site open-closed transition and destabilized the cooperativity between catalytic subunits. Remarkably, mutational changes were always preceded by long periods of accumulation of stabilizing mutations. This highlights the importance of protein flexibility, a reported beneficial trait that is evolutionarily conserved and facilitates quaternary structure assembly [67]. While quaternary structure is conserved in over 70% of protein pairs with <30% sequence identity, in absence of conservation, (dis)assembly pathways mimic evolutionary pathways in processes that resemble Haeckel's biogenetic law [68]. Thus, assembly and evolution appear intimately related. Taking advantage of this link, a 'periodic table' has been constructed that classifies the structure of thousands of homomeric and heteromeric complexes in search for organizing principles [69]. Remarkably, assembly steps can predict likely quaternary structure topologies. This provides a framework to study evolution of quaternary structure. Similar frameworks must be developed for aggregations of polypeptide chains into complex structures integrated with much more complex cellular machinery.

## 2.4. Structural domains are units of protein classification

Because proteins are aggregates of molecular parts with different evolutionary histories, the focus of protein classification for decades has been the structural domain and not the protein molecule. The early recognition that arrangements of secondary structures fell into general protein classes of higher-level structure (originally α/β, α+β, all-α, all-β; [70]) was later formalized into dozens of domain classification schemes that use structure and common evolutionary origin as organizing principles. Two of them, the Structural Classification of Proteins (SCOP) [71] and the CATH database [72] are considered taxonomical gold standards, and a third one, DALI is a structure comparison tool that has been serving the structural biology community for decades [73]. In SCOP, and its recent SCOP2 [74] and extended SCOPe [75] versions, domains that are closely related at sequence level (with pairwise amino acid identities >30%) are grouped into *fold families* (FFs), FFs whose structure and functional features indicate a common evolutionary origin are further grouped into *fold superfamilies* (FSF), FSFs exhibiting tertiary structures with identical topologies are unified into *folds* (Fs), and Fs sharing general architectural designs are finally grouped into 7 *classes* (Table 1). Similarly, CATH uses expert systems to classify domain families into a *homologous superfamilies*, *topologies*, *architectures* and *classes*, with architectures describing the orientation of secondary structures [72]. Finally, the DALI server is a fully automated non-hierarchical alignment system that uses a distance matrix alignment to construct lists of structural neighbors [76]. Remarkably, significant agreement exists between the SCOP, CATH and DALI taxonomies [77].

The cornerstone of these domain hierarchies is common ancestry, i.e. the existence of shared-and-derived features in domain sequence, structure and function. However, the classification of domains does not require structural or functional information nor stringent phylogenetic tests of homology, especially because common ancestry is stronger at lower levels of the classification hierarchy. Most databases benefit from machine learning tools of sequence comparison, including probabilistic hidden Markov model (HMM) methods such as HMMER [78] and HHsearch [79], which conduct HMM-sequence and HMM-HMM alignments, respectively. For example, the Pfam database [80] identifies conserved domain sequences via sequence alignment, which are then used to build HMMs of linear sequence analysis restricting the focus to the sequence level. Conversely, SCOP uses HMMs of structural recognition to recurrently enrich the database [81] in a framework that increases alignment-quality and stability of family and superfamily relationships. Finally, DALI provides structural alignments as either 3-dimensional (3D) or 2-dimensional comparisons by explicitly rotating and translating one domain structure over another or by mapping 3D structure into a matrix of intramolecular distances, respectively [76]. Since structure is far more conserved than sequence, structural similarities can therefore dissect deeper homology relationships than sequences, especially when these are established between domain regions of different sizes.

While not all domains fold into discrete structural entities within the space of possible folds (a fact known as 'gregariousness' [82]), there may be structural similarities at higher levels of classification (e.g. [83,84]). Such statements, however, need to be tested with rigorous phylogenetic approaches.

**Table 1.** Comparison of SCOP, SCOPe, SCOP2, CATH and DALI classification systems of structural domains. Level 1 describes the gross primary secondary structure content of structural domains. Level 2 describes the orientation, connectivity and topology of secondary structures. Level 3 describes structures sharing common ancestors with similar biological function. Level 4 describes structures with sequence similarities of >30 for SCOP, ≥35 for CATH and <25% for DALI (using non-redundant PDB25).

| Level | SCOP 1.75 (June 2009) (hierarchical) | SCOPe 2.07 (January 2021) (hierarchical) | SCOP2 1.0.5 (April 2021) (hierarchical) | CATH 4.3 (July 2019) (hierarchical) | DALI (April 2021) (threshold-delimited) |
|---|---|---|---|---|---|
| 1 | Class (7) | Class (7) | Class (7) | Class (6) | – |
| 2 | Fold (1,195) | Fold (1,232) | Fold (1,500) | Architecture (41) Topology (1,390) | – |
| 3 | Superfamily (1,962) | Superfamily (2,026) | Superfamily (2,681) | Homologous Superfamily (6,631) | – |
| 4 | Family (3,902) | Family (4,919) | Family (5,627) | Family (32,388) | Chains (21,214) |

## 2.5. Domain organization describes the enhanced complexity of multidomain proteins

The arrangement of domains along the sequence of multidomain proteins is referred to as *'domain organization'*, which together with tertiary structure make up the *'domain architecture'* of proteins (Figure 1B). While initial studies determined that more than two-thirds of protein sequences were longer than an average domain length [85,86], a later analysis in 749 genomes demonstrated that only about a third of proteins were multidomain [14]. Domains had mean lengths of 256, 280 and 281 amino acids for proteomes of Archaea, Bacteria and Eukarya, respectively. However, protein lengths in Eukarya almost doubled those in Bacteria and Archaea. Shorter prokaryotic nondomain sequences that link domains to each other in proteins accounted for the difference, suggesting these linkers evolved reductively in prokaryotes but not in eukaryotes [14].

The existence of a modular landscape of domain organization suggests that the recruitment of domains has played a crucial role in protein evolution [87]. The combination of domains appears driven by domain fusions, fissions, circular permutations, insertions and domain loss (e.g. [88,89]). Domains in multiple architectural contexts have been quantified for example by counting distinct domain neighbors [90], adjacencies [91], or consecutive triplets [92]. These measures of 'versatility', 'promiscuity' or 'mobility' (reuse) of domain building blocks depend on both domain size and abundance. Smaller domains are more likely to be used in multidomain proteins and are therefore more mobile [92,93], an observation supported by a Menzerath-Altmann's law of domain organization in which larger molecular systems have smaller parts [94]. Similarly, highly abundant domains appear more versatile. Domain co-occurrence networks have been constructed in which domain nodes are linked when proteins share connected domains [95-98]. In these networks, reticulation indicates evolutionary recruitment. The more tangled the network the less vertical phylogenetic processes drive domain evolution. Note that these networks differ from networks in which domains are linked by a threshold of sequence similarity (e.g. [99]).

## 3. Protein domains distributions in cellular organisms and viruses

The comparative genomic analysis of structural domains in proteomes that began with Gerstein [100] informs about the lexicon, syntax and semantics of proteome vocabularies [101]. For example, Table 2 shows a decade-spanning series of analyses of SCOP FSF and FF distributions in the proteomes of superkingdoms and viruses [102-105]. The existence of 'cores' common to cellular life and viruses (Venn groups ABEV) and cellular life (ABE) (making up 20-26% of all domains) support both a last universal common ancestor (LUCA) and a last universal cellular ancestor (LUCellA) of life. A 'periphery' of domains specific or shared by groups suggest their late diversification, including vocabulary compression typical of microbial life [101]. Significant distribution biases, such as the supernumerary BE and BEV Venn groups, support a tripartite cellular world and an early origin of Archaea [106]. Groups that include viruses, while underrepresented, support wide structural exchange and very early viral origins [104,105]. Similar comparative genomic patterns can be extracted from Pfam domain data [107]. However, universal Pfam cores barely exceed 10%, showcasing the limitations of lower protein organization levels. Percent domain composition of each Venn group was quite constant despite significant increases in proteome and viral sampling (Table 2), suggesting comparative genomic patterns are robust. However, processes of reticulate evolution such as horizontal gene transfer complicate interpretations, prompting phylogenomic analyses.

**Table 2.** Comparative genomic analyses of structural domains across cellular superkingdoms and viruses. SCOP FSF or FF domain distributions define Venn groups shared by superkingdoms and viruses (ABEV, BEV, AEV, AV, BV, EV), superkingdoms (ABE, AB, AE, BE), and superkingdom- or virus-specific (A,B,E,V). Total domain numbers are also given for individual superkingdoms and viruses. Percentages of distinct domains identified in each study are provided in parentheses.

| Venn distribution groups | | Phylogenomic study[1] | | | |
|---|---|---|---|---|---|
| | Dataset: | D1 | D2 | D3 | D4 |
| | SCOP level: | FSF | FSF | FSF | FF |
| | Proteomes: | 1,037 | 4,211 | 5,080 | 8,127 |
| | Viruses: | 56 | 2,715 | 3,460 | 6,044 |
| Total | | 1,739 | 1993 | 1,995 | 3,892 |
| Total Archaea | | 885 (50.9%) | 1014 (50.9%) | 1,022 (51.2%) | 2,198 (56.5%) |
| Total Bacteria | | 1,312 (75.4%) | 1532 (76.9%) | 1,535 (76.9%) | 3,216 (82.6%) |
| Total Eukarya | | 1,508 (86.7%) | 1633 (81.9%) | 1,661 (83.3%) | 3,131 (80.4%) |
| Total Viruses | | 304 (17.5%) | 654 (32.8%) | 716 (35.9%) | 1,526 (39.2%) |
| ABEV | | 229 (13.2%) | 395 (19.8%) | 442 (22.2%) | 979 (25.2%) |
| ABE | | 557 (32.0%) | 492 (24.7%) | 457 (22.9%) | 899 (23.1%) |
| BEV | | 33 (1.9%) | 96 (4.8%) | 111 (5.6%) | 225 (5.8%) |
| ABV | | 2 (0.1%) | 9 (0.4%) | 9 (0.4%) | 52 (1.3%) |
| AEV | | 7 (0.4%) | 8 (0.4%) | 10 (0.5%) | 13 (0.3%) |
| AB | | 36 (2.1%) | 48 (2.4%) | 44 (2.2%) | 168 (4.3%) |
| AE | | 31 (1.8%) | 35 (1.8%) | 32 (1.6%) | 49 (1.3%) |
| BE | | 291 (16.7%) | 283 (14.2%) | 285 (14.3%) | 485 (12.5%) |
| AV | | 1 (0.06%) | 3 (0.2%) | 4 (0.2%) | 5 (0.1%) |
| BV | | 1 (0.06%) | 37 (1.9%) | 33 (1.7%) | 84 (2.2%) |
| EV | | 25 (1.4%) | 39 (2.0%) | 41 (2.1%) | 73 (1.9%) |
| A | | 22 (1.3%) | 24 (1.2%) | 24 (1.2%) | 33 (0.8%) |
| B | | 163 (9.4%) | 172 (8.6%) | 154 (7.7%) | 324 (8.3%) |
| E | | 335 (19.3%) | 285 (14.3%) | 283 (14.2%) | 448 (11.5%) |
| V | | 6 (0.3%) | 67 (3.4%) | 66 (3.3%) | 95 (2.4%) |

[1]D1: [102]; D2: [103]; D3: [104]; D4: [105].

## 4. Protein and proteome evolution

### 4.1. Questions of history are necessary to understand biology

When did the first proteins appear in evolution? What was their makeup? Were they structurally simple or complex? Were they ordered or disordered? What were their biological functions? How were they synthesized? When and how did they start to interact with nucleic acids? When did proteins diversify into modules? How did modularization occur? When did proteins assemble into protein complexes and proteomes? How did these assemblies diversify? What were the evolutionary drivers for all of these processes?

These many questions of deep molecular evolution and others we have not asked can be answered using two general and mutually reinforcing strategies based on either the 'nomothetic' and 'ideographic' scientific methods introduced by Wilhelm Windelband at the end of the 1800s and recently discussed [108]. The nomothetic (universal, predictive) method seeks universal statements that explain individual (preferably present) events with high predictive power. Ideographic (historical, retrodictive) methods support the science of process and history, seeking explanations of how present events have been molded by the past. Both approaches have goals of forecasting-molding events in the future. Origin of life and molecular evolutionary questions can be answered with: (i) 'bottom-up' nomothetic methods that seek prediction of evolutionary likelihoods, and (ii) 'top-down' ideographic methods that seek historical reconstructions from extant data. One drives strength from frequentist or Bayesian views, the other from evolutionary processual and phylogenetic approaches. While both strategies have the potential to reveal how reality is affected by time [109], our focus will be mainly ideographic: the construction of molecular *'chronologies'*, i.e. the arrangement of innovation events (perdurantists's temporal parts) in the order of their temporal occurrence.

In order to retrace past events in protein, proteome and nucleic acid evolution, ideographic statements of history (*phylogenies*) must be proposed directly from genomic data and used to build chronologies of first evolutionary appearance of biological entities such as protein domains and domain architectures or substructures that make up RNA. These chronologies describe the step-by-step history of biological systems and subsystems. Unfortunately, protein and nucleic acid sequence has limited power to dissect deep evolutionary relationships (e.g. [110,111]). Furthermore, while structure is conserved over longer evolutionary timescales, a general metric for global pairwise comparison of protein structures does not yet exist [112]. Thus, protein structural classification cannot unify widely divergent structures at any level of abstraction (e.g. SCOP superfamilies in a 'galaxy' of folds; [113]). This probably stems from the many neighborhoods that exist in sequence space containing different structures and functions [114]. The absence of rules explaining how structures transform into others has also hampered the use of a 'periodic table' of idealized fold representations [115]. The recently reviewed efforts to dissect the evolution of domain architectures [107] were unable to produce stepwise chronological accounts of the evolution of the protein world. A turning point to overcome so many difficulties however occurred when the focus shifted from molecules to substructures and genomes (reviewed in [8,9]). This was possible thanks to the 'genomic revolution' that unfolded at the turn of the century.

### 4.2. Phylogenomic analyses enable the construction of molecular chronologies

At nucleic acid level, Caetano-Anollés [116,117] was first to build rooted phylogenetic trees describing the evolution of molecules or substructures in those molecules directly from structural or statistical features in RNA. Trees were rooted, i.e. a branch holding the ancestor was identified and 'pulled down' to the base of the tree to reorient change in the direction of time (see review of rooting methodologies [118]). Trees of Life (ToLs) were inferred from trees of molecules. Chronologies of accretion were inferred from trees of substructures (ToSs) describing the gradual growth of ancient molecules such as tRNA, RNase P RNA, 5S rRNA and the rRNA (e.g. [119]). At protein level, Gerstein [120] was first to reconstruct trees of proteomes (ToPs) using distance and maximum parsimony (MP) optimality criteria from a proteomic census of presence-absence (occurrence) of structural domains. His trees however were not rooted. Our group later reconstructed rooted ToPs using MP methods from both domain occurrence and abundance [121], providing tools to track the origin and evolution of proteomes. Others also later extended studies to a larger number of proteomes (beginning with [122]) and with automated updates [123]). We also reconstructed a novel type of phylogenomic tree that describes the evolution of proteome parts [121]. These trees of domains (ToD) can be directly used to build domain chronologies.

Figure 2A describes the rationale of building trees of wholes and parts from nucleic acid and protein molecules and their molecular functions. Genomic censuses allow construction of phylogenetic data matrices with rows and columns representing either phylogenetic 'characters' or 'taxa', both of which can be either parts or wholes of the nucleic acid or protein structural system. ToPs and ToDs can be directly derived from these data matrices using appropriate algorithmic and phylogenetic model implementations [118]. Typically, ToPs are well-balanced phylogenies, i.e. they spread novelty throughout ToL lineages without significant bias according to speciation-extinction macroevolutionary models [124]. In contrast, ToDs are highly unbalanced. Their pectinate (comb-like) appearance accommodates stepwise generative (microevolutionary) models of innovation [125]. Chronologies require establishing how old are domains and proteomes, i.e. finding their age. This can be accomplished using two major strategies (Figure 2B). One establishes character state changes in each branch of a ToP (or ToLs) with algorithmic tools of ancestral character state reconstruction (CSR)[126] based on MP, maximum likelihood or Bayesian inference [127]. It produces partial chronologies because CSR cannot trace changes that are older that the common ancestor (proteome) of diversified life. Conversely, the other strategy derives a full chronology of parts directly from the highly unbalanced ToDs by counting nodes from the root of the phylogeny and expressing the time of origin of domains (their age) as a node distance (nd) in a relative 0-1 scale (e.g. [121,128]). The chronology of domains takes full advantage of the powerful serial homologies defined by domain abundance that are typically used in morphological analyses. It starts with the oldest (first) domains and ends with those that are contemporaneous (at nd = 1). Figure 2C shows a data matrix of genomic occurrences of SCOP FSFs depicted as evolutionary heat map. The matrix reveals reductive evolutionary patterns of loss in bacterial and archaeal proteomes unfolding along the chronology of domains. It also reveals how loses are homogeneously distributed in the proteomes of Archaea but heterogeneously apportioned in Bacteria (discussed in [101]).

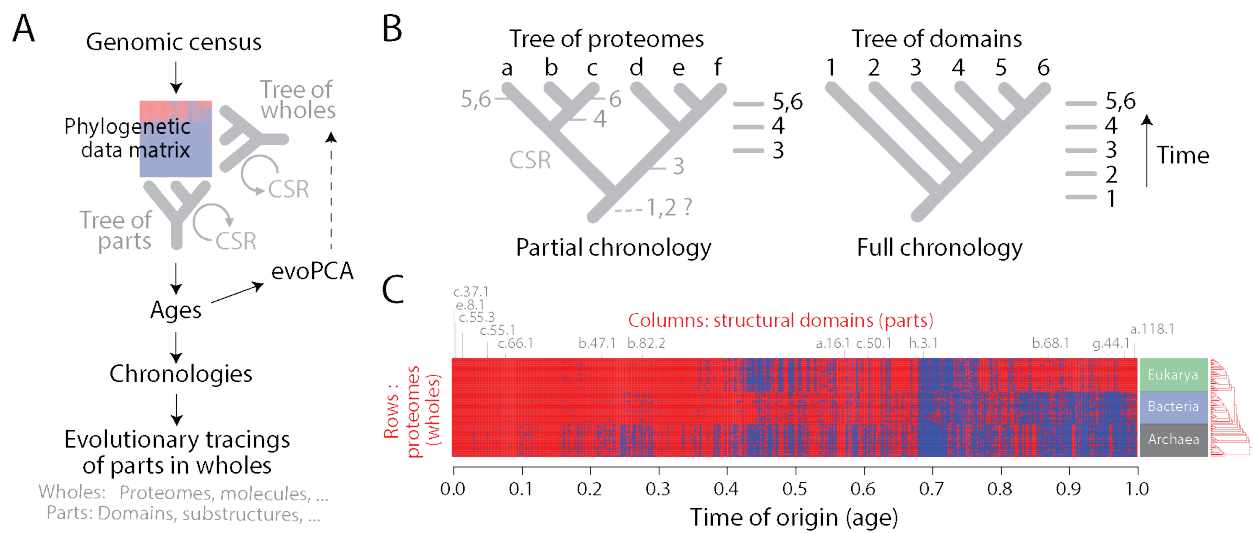

**Figure 2.** Building chronologies with phylogenomics. A. A flow diagram describes the general strategy used to reconstruct trees from genomic data with which to derive times of first appearance (ages) of biological parts and evolutionary chronologies either directly or using ancestral character state reconstruction (CSR) techniques. B. At protein level, chronologies can be derived from trees of proteomes by CSR or directly from trees of domains with a node distance (nd) measuring the relative number of nodes from the base (root) of the phylogeny. CRS provides only a partial chronology delimited by branches arising from a common ancestor. Trees of proteomes allows to derive the time of origin (age in nd units) of each of its leaves (domains). C. An evolutionary heat map describes a phylogenetic data matrix of genomic occurrences. Columns describe 442 universal SCOP FSFs shared with viruses detected in ~11 million proteins ordered according to their time of origin (nd) and rows describing a 368-proteome subset ordered according to a rooted tree of proteomes. FSF presence-absence is described as red-blue cells, respectively. Data from Nasir and Caetano-Anollés [104].

### 4.3. The molecular clock of folds links molecular and geological evolution

Molecular clocks provide powerful frameworks for timescales. We introduced a molecular clock of protein folds that assigns time of origins of domains (nd values) to geological ages derived from fossils and geochemical, biochemical and biomarker data [129]. To showcase one calibration point, the spectroscopic identification of vanadyl-porphyrin complexes in carbonaceous matter embedded in a 3.49 Gy old polycrystalline rock [130] was linked to the age of the periplasmic phosphate-binding protein-like (c.94.1.1) fold family of enzymes of porphyrin biosynthesis, which appeared very early in domain history ($nd_{FF} = 0.018$). Since calibrations resulted in significant linear correlations between geological time and the age of SCOP folds, FSFs and FFs [129,131], the molecular clock was used to establish molecular geochronologies. It showed aerobic metabolism emerged 2.9 Gya and expanded until the Great Oxydation Event (GOE) 2.45 Gya [129], likely triggered by the discovery of the Mn catalase enzyme, which produced oxygen as a side product of $H_2O_2$ detoxification during the Pongola glaciation [131]. Figure 3 (bottom insert) shows the clock behaves well for FFs ($R^2 = 0.900$; $P < 0.001$) using time of origins of domains derived from a recent domain census [106]. The clock falsifies the idea that domains appeared together very early in evolution by assembly from smaller molecular parts and were later lost or gained in lineages. Instead, it supports domain origins that materialized throughout the 3.8 Gy history of protein evolution. This challenges the narrative that gene diversification is solely driven by gene duplication. Instead, domains likely arise step-by-step from different regions of sequence space through structural innovations and recruitments.

## 4.4. Chronologies uncover principles of structural change and molecular function

Biological chronologies dissect evolutionary patterns with which to identify constraints imposed by fold space or establish rules or principles of structural evolution [8]. For example, early work showed an α-to-β structural tendency driving evolutionary change [121]. Similarly, β-barrels with increased curl and stagger were favored evolutionary outcomes in the all-β class. Chronologies of SCOP FFs uncovered the early evolutionary rise of sandwiches and bundles, closely followed by half-barrels and barrels [132], confirming earlier findings at fold and FSF levels (e.g. [121,133]). Table 3 describes a published chronology of CATH architectures [134]. It shows the early appearance of layered 'sandwich' structures and α-β complex arrangements that make popular CATH architectural designs. These architectures are followed by bundles and then barrels, and finally by more complex and symmetric folds, including sheets, rolls and ribbons, solenoid and horseshoes, specialized or irregular structures, and propellers and prisms, in that order. Within each of these broad groups there were also interesting evolutionary patterns. For example, the oldest β-propeller fold contains 6 antiparallel β-sheets in radial arrangement. Later structural innovations increased the number of sheets of the closed structure from 6 to 8, and only then propellers with lower sheet components appeared in evolution. Similarly, the 3-layered α-β-α sandwich evolved into a 3-layered β-β-α, which only then transformed into lower and higher layered structures or formed distorted or inverted (β-α-β) sandwich arrangements. These evolutionary patterns could help develop rules for a periodic table of folds [115]. Table 3 also lists the number of GO functions associated with CATH architectures, showing a clear decrease of functional diversity with time. This suggests an architectural tendency to become more specialized. The hypothesis that the oldest fold designs have molecular functions that use the most chemical biocatalytic mechanisms was also tested and confirmed [135]. Table 4 lists the oldest mechanistic step types of the MACiE v. 3.0 database (now M-CSA [136]). These mechanistic definitions of enzymatic functions that were introduced by the oldest 4 CATH homologous superfamilies accounted for over a third of all mechanistic types, which were also the most popular. In fact, the first fold introduced 6 biocatalytic mechanisms, including electron transfer, which were extensively reused throughout the timeline of CATH superfamilies. These results highlight the functional diversity of the very early structural domains. In this regard, chronologies of GO terms directly derived from trees of molecular functions [137,138] revealed the primacy of binding and catalysis, which were embedded in the first GO terms of molecular functions (discussed in [101]). Individual studies of fold groups support these initial functional roles. For example, structural comparisons of the ubiquitous P-loop NTPases and Rossmann folds suggest they emerged from shared 'theme' sequences and a common β-α-β ancestral loop that bound either catalytic metals or nucleotide ribose moieties [139]. Networks of redox metal-binding domains linked by permissive profile alignments suggested there were at least ten different origins of metal-containing oxidoreductases but also ancient connections that unified $Fe_2S_2$ and heme containing enzymes [140]. Comparsion of the taxonomic and functional breath, structural complexity, and cofactor usage of the triosephosphate isomerase (TIM) barrel fold to other mixed α-β layered structures suggested a facile emergence of the barrel structure and its early metabolic centrality [141]. Remarkably, studies of very old and relatively new metabolic enzymes uncovered a physical (compositional and volumetric) and evolutionary conservation trace in protein structure associated with active sites during the progressive emergence of metabolic activities [142].

## 4.5. Chronologies reveal emergence of molecular and cellular innovations

FF domain structures are generally unambiguously linked to molecular functions and are therefore powerful when studying the emergence and recruitment of novel biological functions [132,143]. They have been adopted by the Molecular Ancestry Network (MANET) database [144] to trace enzyme evolution onto metabolic networks [145]. Figure 3 illustrates the power of biological geochronologies with the origin of ancient domains, the diversification of proteomes, the rise of multicellularity, the metabolic origins of translation, and the evolution of protein folding.

(i) *Domains:* A chronology of bipartite networks that link domains to EFLs (and their projections) showed that the multifunctional α-β-α primordial layered design typical of P-loop and Rossmann-like sandwich structures was primordial [146]. Two waves of domain innovation were jumpstarted by EFL7, EFL6488 and EFL6739 of the nucleotide triphosphate-binding P-loop structure and by EFL2914 (in black and red, respectively; Figure 3A).

**Table 3.** Chronology of CATH architectures. The times of origin of architectures are given in node distances (nd) from the root of the ToD. Architectures are grouped according to general structural designs [1] and indexed wth number of GO functions. Data brom Bukhari and Caetano-Anollés [134].

| CATH A | CATH architecture description | Age (nd) | LC | Bun | Barr | SRR | SH | SI | Prp | Pr | No. Functions |
|---|---|---|---|---|---|---|---|---|---|---|---|
| 3.40 | 3-Layer(aba) Sandwich | 0 | X | | | | | | | | 23 |
| 1.10 | Orthogonal Bundle | 0.034 | | X | | | | | | | 26 |
| 3.90 | Alpha-Beta Complex | 0.069 | X | | | | | | | | 24 |
| 2.40 | Beta Barrel | 0.138 | | | X | | | | | | 27 |
| 3.20 | Alpha-Beta Barrel | 0.138 | | | X | | | | | | 23 |
| 3.50 | 3-Layer(bba) Sandwich | 0.138 | X | | | | | | | | 21 |
| 3.30 | 2-Layer Sandwich | 0.172 | X | | | | | | | | 22 |
| 1.20 | Up-down Bundle | 0.241 | | X | | | | | | | 23 |
| 2.30 | Roll | 0.241 | | | | X | | | | | 24 |
| 3.60 | 4-Layer Sandwich | 0.241 | X | | | | | | | | 12 |
| 2.160 | 3 Solenoid | 0.276 | | | | | X | | | | 10 |
| 2.70 | Distorted Sandwich | 0.310 | X | | | | | | | | 15 |
| 2.60 | Sandwich | 0.345 | X | | | | | | | | 23 |
| 1.25 | Alpha Horseshoe | 0.379 | | | | | X | | | | 25 |
| 4.10 | Irregular | 0.414 | | | | | | X | | | 23 |
| 2.170 | Beta Complex | 0.448 | X | | | | | | | | 18 |
| 2.120 | 6 Propellor | 0.517 | | | | | | | X | | 21 |
| 1.50 | Alpha/alpha barrel | 0.552 | | | X | | | | | | 12 |
| 2.10 | Ribbon | 0.552 | | | | X | | | | | 17 |
| 2.20 | Single Sheet | 0.552 | | | | X | | | | | 22 |
| 2.130 | 7 Propellor | 0.552 | | | | | | | X | | 21 |
| 3.80 | Alpha-Beta Horseshoe | 0.552 | | | | | X | | | | 24 |
| 3.65 | Alpha-beta prism | 0.586 | | | | | | | | X | 8 |
| 3.100 | r-protein L15 | 0.621 | X | | | | | | | | 2 |
| 2.102 | 3-layer Sandwich | 0.690 | X | | | | | | | | 9 |
| 2.140 | 8 Propellor | 0.690 | | | | | | | X | | 12 |
| 3.75 | 5-stranded Propeller | 0.690 | | | | | | | X | | 7 |
| 3.55 | 3-Layer(bab) Sandwich | 0.724 | X | | | | | | | | 2 |
| 2.50 | Clam | 0.759 | | | | | | X | | | 2 |
| 2.150 | 2 Solenoid | 0.793 | | | | | X | | | | 2 |
| 2.115 | 5 Propellor | 0.828 | | | | | | | X | | 1 |
| 2.90 | Orthogonal Prism | 0.862 | | | | | | | | X | 3 |
| 2.100 | Aligned Prism | 0.897 | | | | | | | | X | 4 |
| 2.110 | 4 Propellor | 0.931 | | | | | | | X | | 8 |
| 3.15 | Super Roll | 0.966 | | | | X | | | | | 9 |
| 2.80 | Trefoil | 1 | | | X | | | | | | 17 |
| 3.70 | Box | 1 | | | | | | X | | | 5 |

[1] LC, layered structures and complex arrangements of alpha-beta components; Bun, bundles; Barr, barrels; SRR, sheets, rolls and ribbons; SH, solenoids and horseshoes; SI, specialized or irregular; Prp, propellers; Pr, prisms

One wave describes the rise of 'P-loop' domains typical of the CATH 3.40 [3-layer ($\alpha\beta\alpha$) sandwich] architecture and the other describes 'winged helix' domains typical of the CATH 1.10 (orthogonal bundle) architecture, respectively. These CATH architectures are the oldest of the timeline (Table 3). The evolving network revealed major pathways of EFL recruitment of Rossmann-like folds involving three cysteine-rich EFLs, EFL536, the metal-binding EFL1845, and EFL2524, which are responsible for enzymes of planetary significance such as RubisCO (Figure 1) via the glycine-rich EFL8 prototype and heme-dependent catalases. These pathways support the suggestion that the evolutionary formation of new folds by combination of EFLs was linked to metallic cofactor recruitment [147] and are consistent with recent sequence profile-profile alignment and network analyses of ancient folds [148]. The network also shows EFL recruitment events occurred throughout the 3.8 Gy history of proteins. Thus, the origin of novel domains appears an ongoing process [146].

**Table 4.** Mechanistic step types introduced by the oldest four CATH homologous superfamilies. Step types were described in MACiE v. 3.0. Data from Nath et al. [135].

| Mechanism | First introduction[1] | Age (nd) |
| --- | --- | --- |
| Proton transfer | 3.40.50.300 | 0 |
| Bimolecular nucleophilic addition | 3.40.50.300 | 0 |
| Unimolecular elimination by the conjugate base | 3.40.50.300 | 0 |
| Bimolecular nucleophilic substitution | 3.40.50.300 | 0 |
| Electron transfer | 3.40.50.300 | 0 |
| Intramolecular nucleophilic addition | 3.40.50.300 | 0 |
| Bimolecular elimination | 3.40.50.150 | 0.01 |
| Hydride transfer | 3.40.50.720 | 0.01 |
| Aromatic bimolecular nucleophilic addition | 3.40.50.720 | 0.01 |
| Radical formation | 3.40.50.720 | 0.01 |
| Assisted keto-enol tautomerisation | 3.40.50.720 | 0.01 |
| Radical termination | 3.40.50.720 | 0.01 |
| Aromatic unimolecular elimination by the conjugate base | 3.40.50.720 | 0.01 |
| Redox | 3.40.50.720 | 0.01 |
| Bimolecular homolytic addition | 3.40.50.720 | 0.01 |
| Bimolecular electrophilic addition | 3.40.50.720 | 0.01 |
| Aromatic intramolecular elimination | 3.40.50.720 | 0.01 |
| Colligation | 3.50.50.60 | 0.01 |

[1] CATH superfamily entries: 3.40.50.300, P-loop containing nucleotide triphosphate hydrolases; 3.40.50.150, Vaccinia virus protein VP39; 3.40.50.720, NAD(P)-binding Rossmann-like domain; 3.50.50.60, FAD/NAD(P)-binding domain.

(ii) *Proteomes:* The times of origin of FF domains unique or shared among superkingdoms and viruses define six phases of proteome evolution [105], which we make explicit using dataset D4 of Table 2 (Figure 3B). Phase 0 holds the oldest FFs, which belong to the universal ABEV Venn group. They make up a 'pangenome' responsible for the birth of a communal cellular world. In Phase I, younger FFs shared by cells (ABE) signal a split between stem lines of descent leading to cells and viruses. Phase II describes the initial accumulation of BE FFs and both the rise of a stem line common to Bacteria and Eukarya and reductive loss in primordial ancestors of Archaea. Phase III jumpstarts a diversified world with the first superkingdom-specific FFs (the B group) and the slow emergence of Bacteria, which fully materializes in the next phase. Phase IV embodies the late emergence and concurrent diversification of Archaea, Eukarya and modern viruses made explicit by the appearance of FFs specific to those groups. Virus-specific FFs include capsid and coat folds necessary for viral infections. Phase V describes the final diversification of Eukarya. These patterns of proteome evolution have been consistently recovered since first reported for SCOP folds and FSFs [128,149].

(iii) *Multicellularity:* A previous phylogenomic study identified multicellularity-linked folds arising late in evolution [128]. We plotted the accumulation of 376 FFs (from dataset D4) belonging to these folds along the chronology, indexed with Superfamily's 50 molecular functional categories (Figure 4C). As expected, most functions accumulated during Phase IV and V, once superkingdoms began to diversify. To illustrate, we mapped the FFs of the first category to appear in the chronology, 'cell adhesion', which began with the inception of the integrin domain at the end of Phase I (~3.2 Gya). Integrins are essential low-affinity binding proteins that enable cell-cell and cell-extracellular matrix interactions crucial for multicellularity. Recently, traits associated with multicellularity in cyanobacteria were found to emerge when nitrogen fixation prompted development of the filamentous morphology [150]. The nitrogen fixation markers of the clock (c.92.2.2) suggest this occurred ~2.6 Gya (Figure 4) at the end of phase II and before the rise of multicellularity FFs.

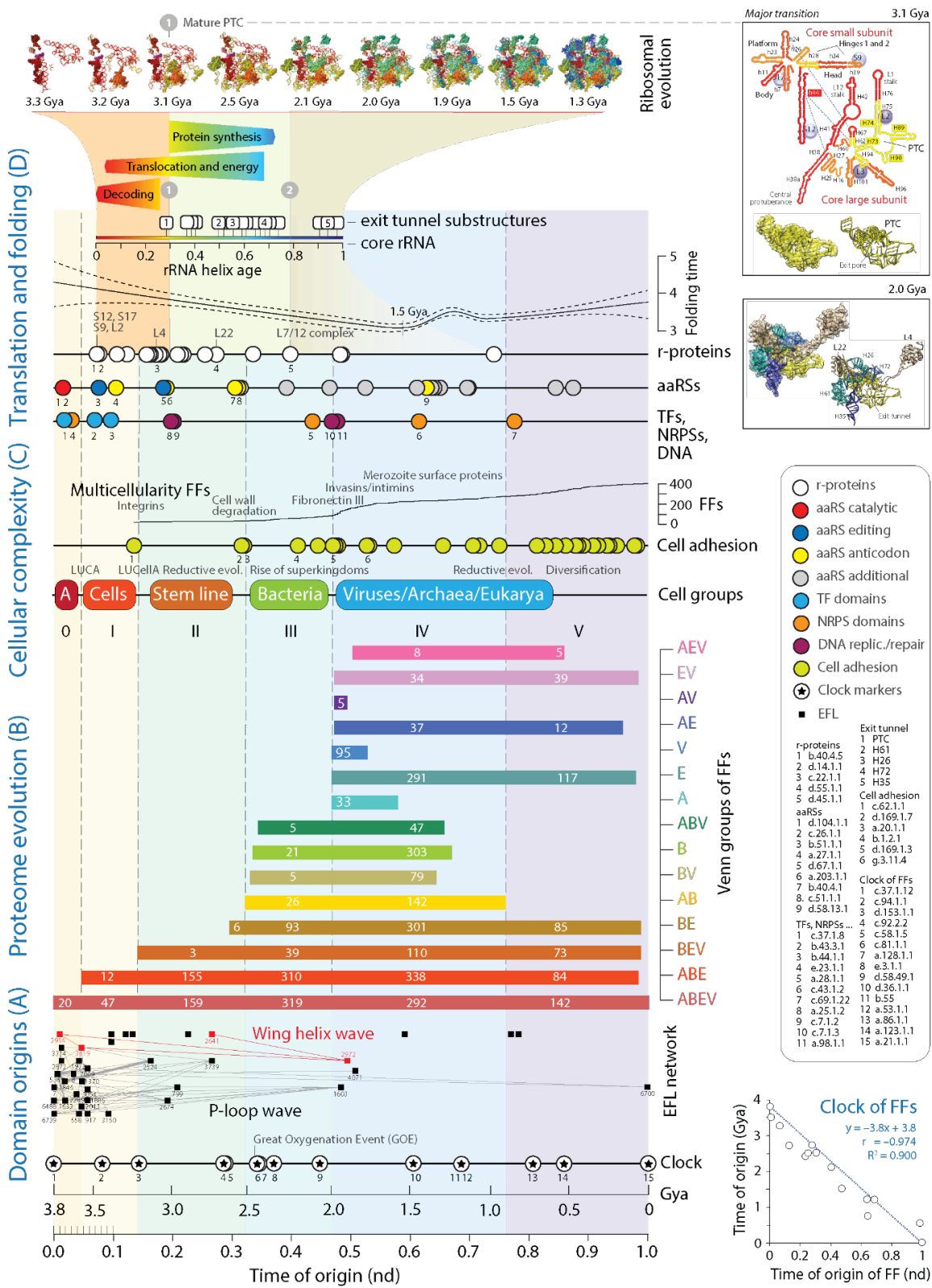

**Figure 3.** Chronologies of elementary functionomes, proteomes, multicellularity, translation and ribosomal structures associated with protein folding. A molecular clock of folds (bottom insert) established that the times of origin (age) of SCOP FF domains measured as node distances (nd) along branches of a ToD were linearly correlated with geological time in billions of years (Gy). Geochronological markers are displayed along the timeline and indexed with numbers. A. Domain origins. The EFL

projection of a bipartite 'elementary functionome' network that links domains to EFLs [146] shows two waves of innovation as EFLs combine to form novel domain functions. B. Proteome diversification. Bar plots show ranges of times of origin (nd) for FFs unique or shared among Archaea (A), Bacteria (B) and Eukarya (E) and viruses (V), as time progresses from the origin of domains (nd = 0) to the present (nd = 1). Numbers in bars indicate FFs appearing in each evolutionary phase of the timeline. C. Cell complexity. Cell groups describe a most likely chronology inferred from Venn group distributions in which ancestral cells (A) coalesce into a last universal common ancestor (LUCA), which then diversifies into a last universal cellular ancestor (LUCellA) and then modern cells. The plot describes FF accumulation of domains linked with multicellularity, and the chronology describes FFs with 'cell adhesion' functions. D. Translation and folding. Chronologies of FFs of universal r-proteins, aaRSs, TFs, NRPSs and enzymes linked to deoxyribonuceotide synthesis are mapped to the origin and evolution of the ribosome (structural timeline in the top from [119]. A chronology of rRNA substructures unfold gradual development of ribosomal functions and construction of the exit tunnel once the PTC was formed in a 'major transition' in ribosomal evolution (labeled with an encircled '1'). The plot describes folding time of FFs assessed with size-modified contact order [27] and a chronology of 26 rRNA structures subtending (15 Å-away) the exit tunnel. The inset describes a model of ribosomal secondary structures present during the 'major transition' that occurred 3.1 Gya and a radial vestibular 3D view of the junction that subtends the peptidyl transferase center (PTC) that was formed 3.1 Gya and the exit tunnel formed 2 Gya (described with r-proteins L4 and L22 and with 5 indexed tunnel rRNA substructures), showing its late evolutionary appearance. Note that L4 binds to H26 and L22 binds to H73 of the PTC, H61 and H26, showing the crucial role of these r-proteins in tunnel formation. FF ages in panels B-D are derived from dataset D4 (Table 2) from Mughal et al. [105]. When indexed, FFs are given as SCOP concise classification strings.

(iv) *Translation:* FFs associated with translation and its protein biosynthetic apparatus appeared late in evolution [132,143], following the explosive rise of metabolic enzymes [133]. This once again challenged the validity of the ancient 'RNA world' hypothesis. The first translation domains had catalytic functions for the aminoacylation and the molecular switch-driven transport of RNA. Figure 3D shows that FFs of dataset D4 making up aminoacyl-tRNA synthetases (aaRSs), translation factors (TFs), NRPS modules, and r-proteins appeared in that order during Phases 0 and I. Catalytic domains of aaRSs were first complemented with editing domains, and then with anticodon-binding and auxiliary domains, following a previously described temporal sequence of emergence associated with genetic code specificities [151]. FF domains responsible for ribonucleotide reductase activity (c.7.1.3 and a.98.1.1) needed for the synthesis of deoxyribonucleotides appeared at the start of Phase IV and the diversified cellular world. This demonstrates proteins and RNA preceded DNA.

(v) *Protein folding and ribosomal evolution:* Expedient folding, which varies from microseconds to hours, must be evolutonarily optimized to favor both activity and stability and shield against unwanted aggregation or degradation of proteins. Folding speed was traced along the chronology of FFs by measuring the size-modified contact order of protein structure, a flexibility-correlated estimation of non-locality of amino acid contacts [27]. Remarkably, a biphasic behavior of folding time was recovered, first steadily decreasing between 3.8 and 1.5 Gy and then increasing until the present [27]. This biphasic pattern paraphrased a similar biphasic pattern in the size of protein domains. Figure 3D shows how this initial evolutionary optimization for rapid folding coincided with the gradual development of the translation machinery and the ribosome. ToSs and ToDs showed r-proteins appeared late in evolution and coevolved with tRNA and the growing rRNA scaffold of the universal ribosomal core [9,119]. In this accretion process, a functioning modern ribosome was assembled after a first ribosomal 'major transition' brought translocation ('turnstile') and polypeptide biosynthetic functions together (Figure 3D, insert). This coordinated event occurred 3.1 Gya, well after the establishment of a protein world, suggesting that the early-appearing NRPSs were already synthesizing proteins in absence of RNA. The universal core continued to accrete until 1.3 Gya. We used dataset D4 (Table 2) to reveal a statistically significant association between the ages of r-proteins and rRNAs [Spearman's rho = 0.566; $p$ (2-tailed) = 0.0006], which allowed to map the ages of associated FFs to the timeline of rRNA accretion and the molecular clock (Figure 3D). Mapping the substructures subtending (15 Å-away) the ribosomal exit tunnel growing from the exit pore of the peptidyl transferase center (PTC) allowed to establish a timeline for co-translation folding, revealing gradual rRNA accumulation during Phases II and III 3.1-to-2.0 Gya. This mapping revealed that the exit tunnel constrictions started to materialize ~2.6 Gya with r-protein L22, which together with L4 define the start of the lower tunnel. Later addition of the late-developing Bacteria- and Eukarya/Archaea-specific rRNA substructures H72 and H35 likely resulted in the observed width differential of the tunnel [35]. This occurred before and after the two domains of r-protein L7/12 that anchor to L10, bind to TFs, and are responsible for the second ribosomal transition that increases ribosomal processivity ~2.2 Gya. The development of the exit tunnel before the transition of folding speed 1.5 Gya suggests vestibular co-translation folding played a crucial role in the folding history of proteins, likely speeding up the process.

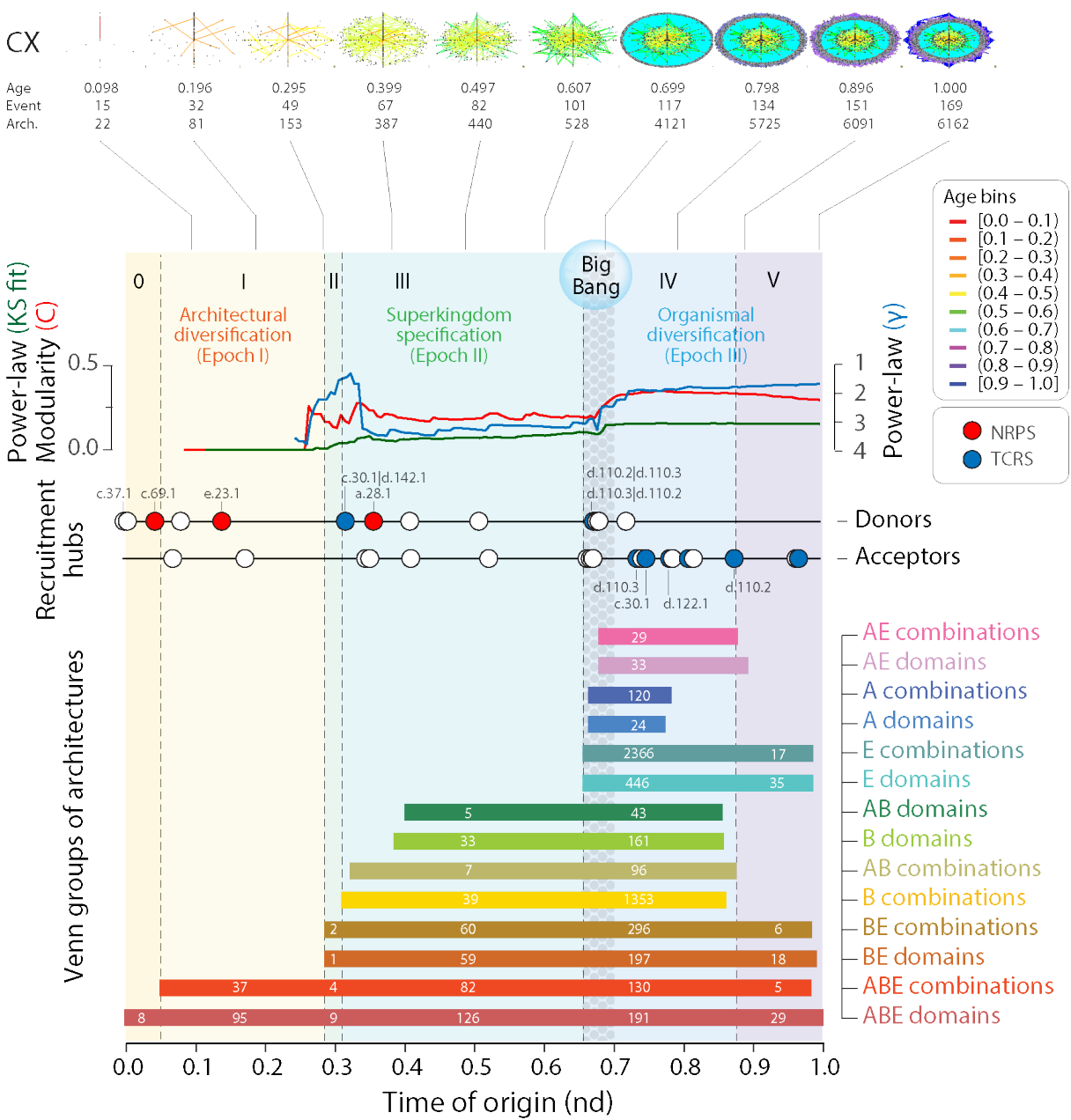

**Figure 4.** Evolution of domain organization visualized with evolving networks and chronologies of architectures grouped according to their distribution in superkingdoms. The time series of networks at the top shows the evolving composition network (CX) sampled at selected time events. CX links domains and supradomains to multidomain nodes when protein share domain makeup at SCOP FSF level. The statistical evaluation of power law and modularity show these are emergent properties of the networks. A chronology of recruitment hubs identifies major recruitment donors and acceptors, with NRPS modules and two-component regulatory systems (TCRS) highlighted with colors. Bar plots show ranges of times of origin (nd) for architectures unique or shared among Archaea (A), Bacteria (B) and Eukarya (E), as time progresses from the origin of architectures (nd = 0), initially only single-domain proteins, to the present (nd = 1). Numbers in bars indicate architectures appearing in each evolutionary phase of the timeline. Data from [98].

## 4.6. Evolutionary 'big bangs' can be visualized with networks

Chronologies describe detailed patterns of growth of molecular diversity in the protein world. This growth sometimes occurs massively (in 'big bangs') such as in the case of biocatalysis [135], metabolism [133], or domain organization [5]. Figure 4 summarizes a recent evolutionary study of protein domain recruitment that uses time series of networks to unfold a chronology of domain organization [98]. Time series and chronologies of architectural accumulation showed massive recruitment responsible for a 'big bang' of domain combinations occurring ~1.5 Gya, a time that coincides with the rise of superkingdoms, multicellularity and molecular innovations responsible for genomic rearrangements. Scale-freeness and modularity were emergent properties of the evolving networks, revealing processes driving network evolution engender functional modules. Major recruitment hubs included cofactor-supporting structures of NRPSs, likely crucial to the development of the genetic code, and two-component regulatory systems (TCRS) necessary for transcriptional regulation of responses to environmental change. These analyses unfold historical source-sink relationships in the recruitment of architectures and function. Remarkably, mapping the times of origin of FF domains unique or shared among superkingdoms once again revealed the six phases of proteome evolution uncovered in Figure 3B, confirming again important patterns of proteome evolution.

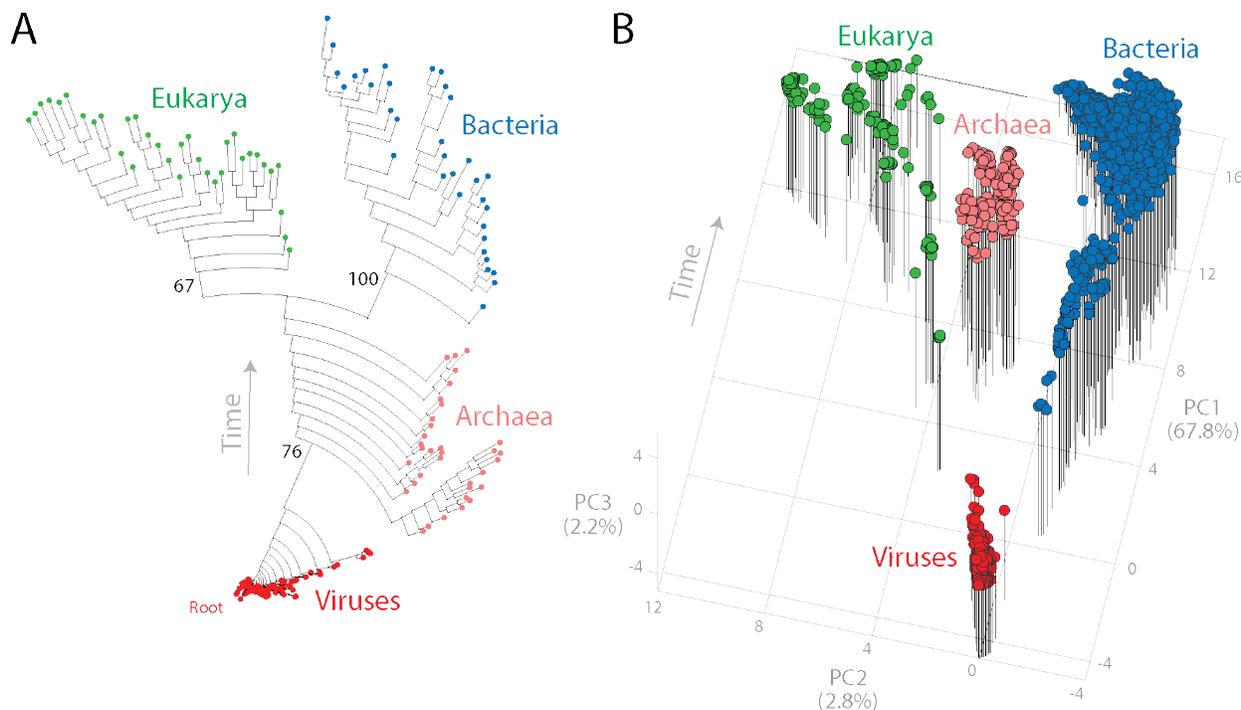

**Figure 5.** Trees of life derived from genomic censuses of structural domains in proteomes. A. ToP reconstructed from 136 proteomes randomly sampled from FSF-based dataset D3 (34 for each supergroup) rooted using the Lundberg method. B. Evolutionary principal coordinate (evoPCO) analysis describing the evolution of 8,127 reference-quality proteomes of cellular organisms and viruses belonging to dataset D4. The 3-dimensional scatter plot describes the temporal relationships of 139 archaeal, 1,740 bacterial, 210 eukaryal and 6,044 viral proteomes using the the ages of 3,892 FFs. The multi- dimensional scaling analysis highlights in its first three most significant axes a temporal flow from viruses to cellular organisms.

### 4.7. Proteome evolution can be visualized through the history of their protein components

ToPs are direct statement of proteome history. Figure 5A for example describes a ToP reconstructed from proteomes randomly sampled from dataset D3 (Table 5). The tree places viruses at its root and Bacteria and Eukarya as derived sister groups [104], matching the progression of Venn groups of Figure 3 and 4 and supporting an Archaea-first scenario for the diversification of cellular life [106]. Figure 5B confirms this history of organismal diversification by exploring the collective history of FF domains in proteomes using dataset D4. A multidimensional scaling strategy termed evolutionary principal coordinate (evoPCO) analysis [104] places proteomes within a multidimensional landscape of times of origin of FFs, promising to overcome some of the limitations of ToL reconstruction methods, including violations of phylogenetic character independence. The first three coordinates account for 72% of total variance and reveal four distinct temporal clouds of proteomes corresponding to the four supergroups of life and a temporal flow from viruses to cellular organisms unfolding along the PCO1 axis. Results reinforce the notion of an ancient origin of RNA viruses, the early diversification of Bacteria revealed in Figures 3 and 4, and a tripartite cellular world (discussed in [106]).

**Table 5.** The time of origin (age) of SCOP FF domain landmarks associated with the metabolic origins of translation. The age of domains (nd) was inferred from phylogenomic analyses of 8,127 proteomes from cellular organisms and viruses (dataset D4; [105] and from 420 free-living organisms (dataset FL420; [143]) shows the same evolutionary progression.

| Landmark | FF css | FF domain name | $nd_{FF}$ (D4) | $nd_{FF}$ (FL420) |
| --- | --- | --- | --- | --- |
| ATPases | c.37.1.12 | ABC transporter ATPase domain-like | 0 | 0 |
| Enzymes | c.2.1.2 | Tyr-dependent oxidoreductases | 0.003 | 0.008 |
| Translation factors | c.37.1.8 | G proteins | 0.015 | 0.02 |
| aaRSs | c.26.1.1 | Class I aaRS, catalytic domain | 0.018 | 0.02 |
| aaRSs | d.104.1.1 | Class II aaRS-like, catalytic domain | 0.018 | 0.025 |
| Enzymatic hinges | c.94.1.1 | Phosphate-binding protein-like | 0.018 | 0.032 |
| NRPSs | e.23.1.1 | Acetyl-CoA synthetase-like | 0.029 | 0.045 |
| Translation factors | b.43.3.1 | Elongation factors | 0.069 | 0.073 |
| r-proteins | b.40.4.5 | Cold shock DNA-binding domain-like | 0.072 | 0.114 |
| r-proteins | d.14.1.1 | Translation machinery components | 0.077 | 0.127 |
| Translation factors | b.44.1.1 | EF-Tu/eEF-1/eIF2-$\gamma$ C-terminal domain | 0.095 | 0.114 |
| Ribosomal PTC | d.66.1.2 | Ribosomal protein S4 | 0.215 | 0.253 |

## *4.8. A note on the difficulties of retrodiction*

Reconstructing the past from information in macromolecules requires data, models, and ground plans, all of which can be limiting. While phylogenomic approaches are data-driven, genomic datasets can be imbalanced because of limited survey (unexplored biological 'dark matter'), known overrepresentations (e.g. bacterial pathogens), or hyperdiverse and/or poorly understood organismal groups (e.g. basal eukaryotes) [106, 108, 118]. The problem of 'holobionts' and the species concept can compromise the definition of taxa. Hierarchical structures in biology can compromise the definition of characters. Underlying evolutionary models can be unrealistic (e.g. the problem of 'gaps' in sequence analysis [106]), simplistic (e.g. the problem of structure in sequence analysis [108]), or overfitting (e.g. neglecting model-to-data fitness [152]). Phylogenies may not be appropriate evolutionary ground plans at temporal (e.g. the problem of defining stem lines or lineages) or spatiotemporal (e.g. dissecting endosymbiont evolution) levels [108]. The entire exercise of retrodiction begs better algorithmic implementations that can accommodate increasing taxa, diverse and more numerous characters, and improved visualization.

## **Expert Opinion**

Chronologies and time-varying networks consistently recovered significant patterns of evolutionary progression, some illustrated in Figures 3-5. The relative ages of landmarks have been congruently recovered at different protein domain abstractions for over a decade, showcasing methodological robustness (Table 5). The gradual succession of events tells a story that is straightforward and often unambiguous. For example, the sharing of FFs between superkingdoms and viruses, when displayed along a timeline, unambiguously describes the gradual rise of cellular complexity (Figure 3). Similarly, tracing FFs associated with multicellularity revealed the accumulation of molecular novelties fostering intercellular interactions. A clear picture of the origin and evolution of proteins emerges from phylogenomic reconstruction and a growing knowledgebase. It illustrates novelty accumulation at different levels, from loops to domains and multidomain arrangements in proteins to entire proteome and functionome complements. Models for the origin of protein biosynthesis, the translation machinery, and the ribosome [119,132,151] continue to be supported by new expanded data exploration. Two processes, folding and recruitment appear central to the evolutionary progression. The former increases protein persistence while the later fosters diversity. Chronologically, protein evolution mirrors folding by combining supersecondary structures into domains, developing complex translation machinery to facilitate folding speed and stability, and enhancing structural complexity by establishing long-distance interactions that create novel structural and architectural designs. This recapitulation theory also matches a model of molecular entanglement that we recently proposed [109].

Historical accounts of molecular evolution that are embodied in chronologies and evolving networks extend the limits of what is possible. Time series of events and combination of events provide unanticipated evolutionary descriptions that will help develop predictive tools for applications in biology and medicine. For example, chronologies reveal evolutionary design principles that can be used in protein bioengineering. Similarly, temporal patterns can help define parameters in evolutionary modeling and prediction studies. Phylogenomic analyses promise to unveil evolutionary mechanisms and rationales driving the history of networks of protein domain organization. In addition, time-varying networks are excellent models of dynamical systems [153]. Linking networks and chronologies thus expands the historical account.

Novel phylogenomic methods are still needed to advance the science of deep history reconstruction. Phylogenetic trees lack reticulations that can account for pervasive horizontal exchange processes such as the recruitment of EFLs in domains or domains in proteins. Similarly, phylogenetic networks that could model such recruitments cannot be recovered with optimality criteria because of computational and algorithmic limitations. We therefore trust tree reconstructions can accurately extract vertical evolutionary signatures in a 'sea' of entanglement. Not all phylogenetic characters are suitable either. Standard alignment-dependent methods that for example focus on sequences are powerful for establishing shallow phylogenetic relationships but fail to cleanly recover deep evolution [106]. Alignment-free methodologies overcome some of the challenges but depend on the accuracy of machine-learning methodologies or the levels of evolutionary conservation embedded for example in the hierarchy of loops or domain taxonomies. Finally, networks that exploit sequence or structure similarity lack an homology-based phylogenetic rationale and must be carefully validated with other phylogenetic approaches. Learning how to interface levels of protein structural complexity and tree and network reconstruction will remain an important task for the future.

## Funding

Supported by grants from the National Institute of Food and Agriculture of the United States Department of Agriculture (ILLU-802-909 and ILLU-483-625) and allocations from the National Center for Supercomputer Applications (NCSA).

## Glossary

| | |
|---|---|
| aaRSs | Aminoacyl-tRNA synthetases |
| FF | Fold family |
| FSF | Fold superfamily |
| Gya | Billion years (Gy) ago |
| HMM | Hidden Markov model |
| NRPS | Non-ribosomal protein synthetase |
| TCRS | Two-component regulatory system |
| TFs | Translation factors |
| ToD | Tree of domains |
| ToM | Tree of molecules |
| ToL | Tree of life |
| ToP | Tree of proteomes |
| ToS | Tree of substructures |